\documentclass[aps,prl,floatfix,superscriptaddress,twocolumn,10pt]{revtex4-1}
\usepackage{graphicx}
\usepackage{dcolumn}
\usepackage{bm}
\usepackage{color}
\usepackage{amsmath,amsfonts,amssymb}
\usepackage{graphicx}
\usepackage[english]{babel}
\usepackage{color}
\usepackage{booktabs,longtable}
\usepackage{natbib}
\usepackage{textgreek} 
\usepackage{upgreek} 
\usepackage{mathrsfs}

\usepackage[colorlinks]{hyperref}
\hypersetup{
                pdfstartview={FitH},
                linkcolor=blue,
                citecolor=blue,
                filecolor=blue,
                urlcolor=blue
}

\newcommand\varpm{\mathbin{\vcenter{\hbox{
  \oalign{\hfil$\scriptstyle+$\hfil\cr
          \noalign{\kern-.3ex}
          $\scriptscriptstyle({-})$\cr}
}}}}
\newcommand\varmp{\mathbin{\vcenter{\hbox{
  \oalign{\hfil$\scriptscriptstyle-$\hfil\cr
          \noalign{\kern-.3ex}
          $\scriptstyle({+})$\cr}
}}}}

\begin{document}

\begin{sloppypar}

\title{Observation of Large Unidirectional Rashba Magnetoresistance in Ge(111)}

\author{T. Guillet}
\affiliation{Univ. Grenoble Alpes, CEA, CNRS, Grenoble INP, IRIG-SPINTEC, 38000 Grenoble, France}
\author{C. Zucchetti}
\affiliation{LNESS-Dipartimento di Fisica, Politecnico di Milano, Piazza Leonardo da Vinci 32, 20133 Milano, Italy}
\author{Q. Barbedienne}
\affiliation{Unit\'e Mixte de Physique, CNRS, Thales, Univ. Paris-Sud, Universit\'e Paris-Saclay, 91767, Palaiseau, France}
\author{A. Marty}
\affiliation{Univ. Grenoble Alpes, CEA, CNRS, Grenoble INP, IRIG-SPINTEC, 38000 Grenoble, France}
\author{G. Isella}
\affiliation{LNESS-Dipartimento di Fisica, Politecnico di Milano, Piazza Leonardo da Vinci 32, 20133 Milano, Italy}
\author{~\text{L. Cagnon}}
\affiliation{Univ. Grenoble Alpes, CNRS, Grenoble INP, Institut NEEL, 38000 Grenoble, France}
\author{C. Vergnaud}
\affiliation{Univ. Grenoble Alpes, CEA, CNRS, Grenoble INP, IRIG-SPINTEC, 38000 Grenoble, France}
\author{N. Reyren}
\affiliation{Unit\'e Mixte de Physique, CNRS, Thales, Univ. Paris-Sud, Universit\'e Paris-Saclay, 91767, Palaiseau, France}
\author{J.-M. George}
\affiliation{Unit\'e Mixte de Physique, CNRS, Thales, Univ. Paris-Sud, Universit\'e Paris-Saclay, 91767, Palaiseau, France}
\author{A. Fert}
\affiliation{Unit\'e Mixte de Physique, CNRS, Thales, Univ. Paris-Sud, Universit\'e Paris-Saclay, 91767, Palaiseau, France}
\author{M. Jamet}
\affiliation{Univ. Grenoble Alpes, CEA, CNRS, Grenoble INP, IRIG-SPINTEC, 38000 Grenoble, France}

\date{\today}

\begin{abstract}

Relating magnetotransport properties to specific spin textures at surfaces or interfaces is an intense field of research nowadays. Here, we investigate the variation of the electrical resistance of Ge(111) grown epitaxially on semi-insulating Si(111) under the application of an external magnetic field. 
We find a magnetoresistance term which is linear in current density $j$ and magnetic field $B$,
 hence odd in $j$ and $B$, corresponding to a unidirectional magnetoresistance. 
At ${15~\textup{K}}$, for ${I=10~\text{\textmu A}}$ (or $j=0.33~\textup{A}\,\textup{m}^{-1})$ and ${B=1~\textup{T}}$, it represents $0.5\%$ of the zero field resistance, a much higher value compared to previous reports on unidirectional magnetoresistance. We ascribe the origin of this magnetoresistance to the interplay between the externally applied magnetic field and the current-induced pseudo-magnetic field in the spin-splitted subsurface states of Ge(111). This unidirectional magnetoresistance is independent of the current direction with respect to the Ge crystal axes. It progressively vanishes, either using a negative gate voltage due to carrier activation into the bulk (without spin-splitted bands), or by increasing the temperature due to the Rashba energy splitting of the subsurface states lower than $\sim58\,k_B$. The highly developed technologies on semiconductor platforms would allow the rapid optimization of devices based on this phenomenon.

\end{abstract}

\maketitle

After decades of studies, spintronics has driven its most successful industrial revolutions in the read-out head of magnetic hard disk and in magnetic random access memory \cite{Akerman2005}. In both cases, a long-range magnetic order is the ultimate ingredient, since these applications rely on the giant magnetoresistance (GMR) effect \cite{Baibich1988,Binasch1989}. Due to the seek of magnetic ordering, the investigation of GMR has been wide and successful in ferromagnetic-based layers but still rare in semiconductors. Since a connection with magnetism in semiconductors would be desirable, lots of interest has been devoted to doping semiconductors with magnetic impurities \cite{Ohno1998,Dietl2000}. The low solubility of magnetic ions \cite{Akerman2005} and Curie temperatures below ${200~\textup{K}}$ \cite{Chen2011} still limit the applicability of this technology. An alternative to long-range magnetic order in semiconductors comes from spin-orbit coupling (SOC), the main core of the so-called spin-orbitronics field in semiconducting films and in topological insulators \cite{Chappert2007,Manchon2015}.

Within this field, the investigation of magnetoresistance has recently moved over the standard ferromagnet-related effects \cite{Locatelli2014,Kim2016,Latella2017,Qiu2018,He2017,Zhang2018,He2018}, and a promising new type of magnetoresistance has been observed in the topological insulator Bi$_{2}$Se$_{3}$ \cite{He2017}, and in the two-dimensional electron gas at the SrTiO$_3$(111) surface \cite{He2018}. Since, in both cases, no magnetic order is present, the effect has been related to the characteristic spin-momentum locking \cite{He2017,Zhang2018,He2018}. The detected magnetoresistance exhibits two characteristic features: it is unidirectional (i.e. odd) and linear with the applied magnetic field and electrical current, therefore it has been classified among the \emph{unidirectional magnetoresistances} (UMRs)\cite{He2017,Zhang2018,He2018}. Despite the same angular dependence, this SOC-related UMR has a different origin compared to another type of UMR recently investigated and involving a ferromagnetic layer \cite{Olejnik2015,Avci2015,Yasuda2016,Lv2018}. 

Here, we report the observation of UMR in Ge(111). We ascribe its origin to the Rashba SOC, which generates spin-momentum locking inside the subsurface states of Ge(111). Their presence and spin-texture have already been demonstrated exploiting angle and spin-resolved photoemission spectroscopy \cite{Ohtsubo2010,Ohtsubo2013,Aruga2015,Yaji2015}. Experimentally, we find that the UMR in the Ge(111) subsurface states is drastically larger compared to previous reports \cite{He2017,He2018}. We detect a maximum UMR value equivalent to $0.5\%$ of the zero field resistance, when a magnetic field of ${1~\text{T}}$ and a current of ${10~\text{\textmu A}}$ are applied at ${15~\textup{K}}$. The effect progressively vanishes when increasing the temperature or applying a negative gate voltage due to carrier activation in the bulk valence bands of Ge and to the low value of the Rashba spin-orbit coupling ($\sim58\,k_B$)\cite{Ohtsubo2010}.

\begin{figure}[t!]
\includegraphics[width=0.48\textwidth]{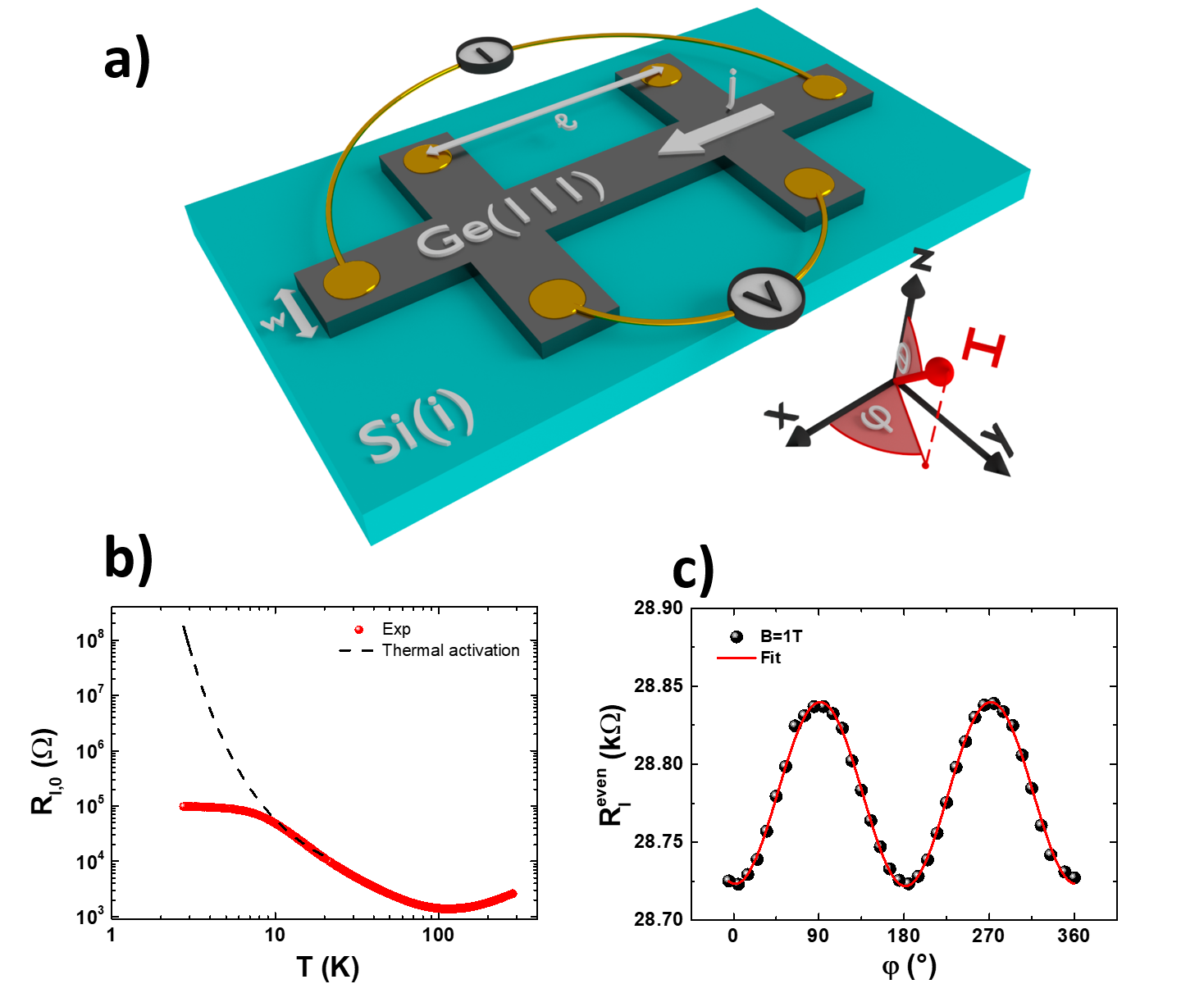} 
\caption{(color online) Sketch of the double Hall cross used for magnetoresistance 4-probe measurements. The external magnetic field is applied along (${\theta,\varphi}$) directions, $\theta$  and $\varphi$ being the polar and azimuth angles. The current is applied along the $[1\bar{1}0]$ crystal axis of Ge. (b) 4-probe resistance versus temperature measured with an applied current of ${10~\text{\textmu A}}$. The red curve corresponds to experimental data exhibiting a resistance saturation, the dashed black line shows the expected semiconducting behavior considering a thermal activation of 2.6 meV. (c) Angular dependence in the $(xy)$ plane of $R_{l}^{\,\textup{even}}$  at ${15~\textup{K}}$. The applied magnetic field is 1 Tesla and the current is ${10~\text{\textmu A}}$. The solid red line is a fit to the experimental data using a sine function.}
\label{Fig1}
\end{figure}

\begin{figure*}[t!]
\begin{center}
\includegraphics[width=\textwidth]{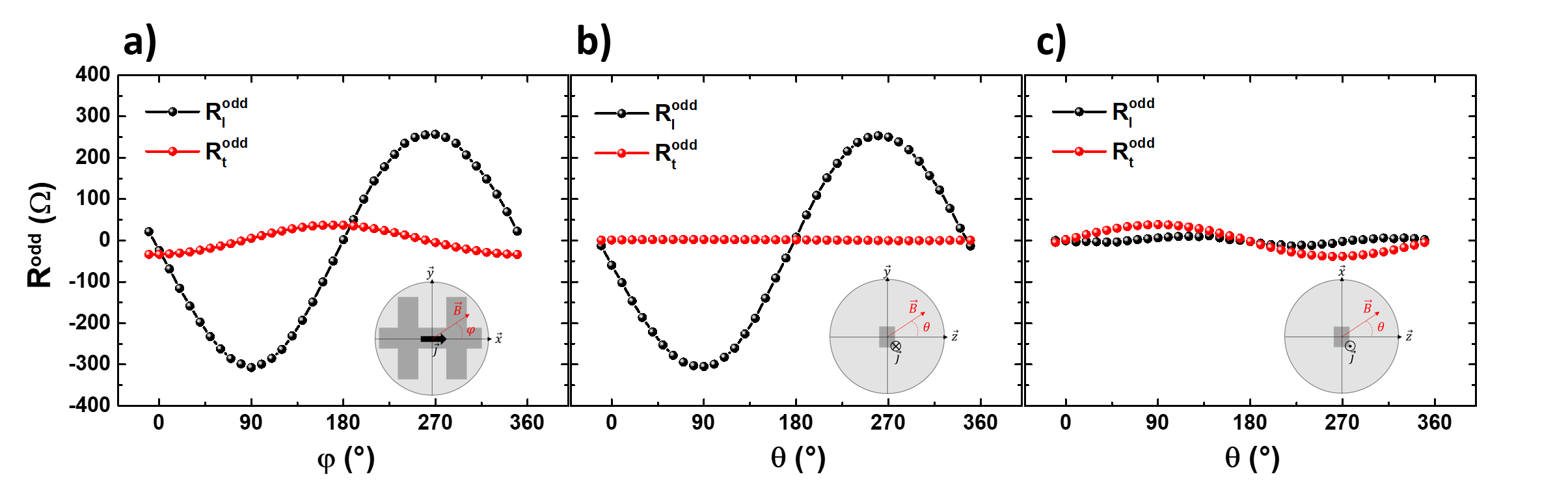} 
\caption{(color online) Angular dependence of $R_{l}^{\,\textup{odd}}$ (black dots) and $R_{t}^{\,\textup{odd}}$ (red dots) in: (a) the ($xy$) plane (${\theta=90^{\circ}}$), (b) the ($yz$) plane (${\varphi=90^{\circ}}$) and (c) the ($xz$) plane (${\varphi=0^{\circ}}$) respectively. The temperature is ${15~\textup{K}}$, the applied current ${10~\text{\textmu A}}$ and the magnetic field ${1~\textup{T}}$. The solid black and red lines are fitting curves of $R_{l}^{\,\textup{odd}}$ and $R_{t}^{\,\textup{odd}}$ respectively using sine and cosine functions. They superimpose to the experimental data.}
\label{Fig2}
\end{center}
\end{figure*}

We perform magnetotransport measurements on a $2~\text{\textmu m}$-thick Ge(111) using lithographically defined Hall bars (length ${\ell=120~\text{\textmu m}}$, width ${w=30~\text{\textmu m}}$ and aspect ratio $Z=\ell /w=4$) as shown in Fig.~\ref{Fig1}\textcolor{blue}{(a)} (further details in the Supplementary Material). We apply a DC charge current and measure the longitudinal ($R_{l}=U_{l}/I$) and transverse ($R_{t}=U_{t}/I$) resistances under the application of an external magnetic field $\textbf{B}$. The direction of $\textbf{B}$ is determined by its polar ($\theta$) and azimuth ($\varphi$) angles as shown in Fig.~\ref{Fig1}\textcolor{blue}{(a)}. In DC measurements, the UMR term is odd with respect to the applied current and thus defined as: ${R_{l}^{\,\textup{odd}}=\,[R_{l}(I)-R_{l}(-I)]/2}$. We also measure ${R_{t}^{\,\textup{odd}}=\,[R_{t}(I)-R_{t}(-I)]/2}$ and the longitudinal resistance which is even with respect to the applied current ${R_{l}^{\,\textup{even}}=\,[R_{l}(I)+R_{l}(-I)]/2}$. All the measurements are carried out as a function of the temperature from ${15~\text{K}}$ to ${295~\text{K}}$. The conductivity is $p$-type in the whole temperature range, at  ${15~\textup{K}}$ the carrier density reaches ${p\approx6\times10^{15}~\text{cm}^{-3}}$.

\begin{figure}[t!]
\begin{center}
\includegraphics[width=0.5\textwidth]{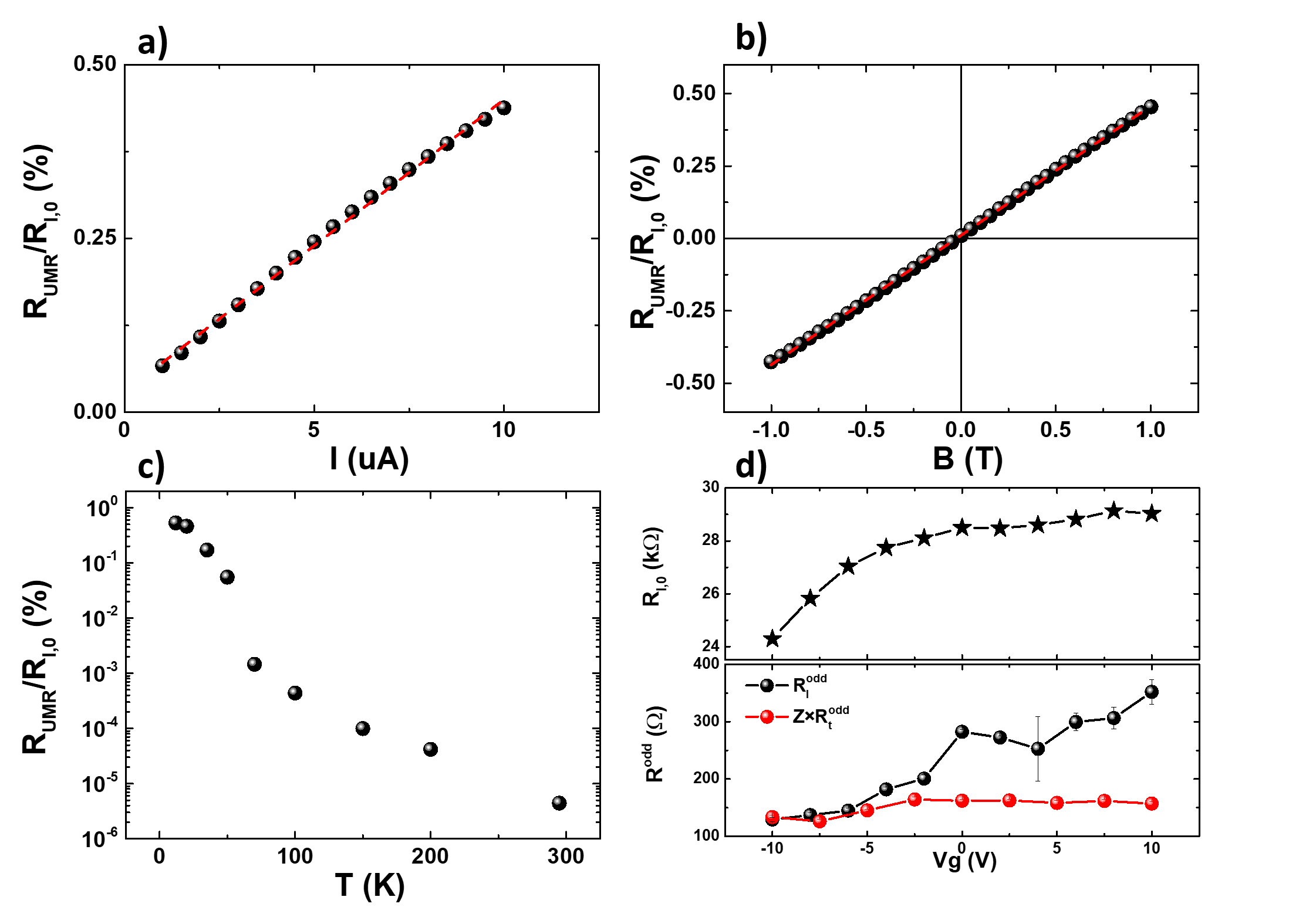} 
\caption{(color online) $R_{\textup{UMR}}$ normalized to the zero magnetic field resistance $R_{l,0}$ taken at $\varphi=~270^{\circ}$ (in \%) as a function of (a) the applied current for ${B=1~\text{T}}$ and ${T=15~\textup{K}}$ (b) the magnetic field for ${I=10~\text{\textmu A}}$ and ${T=15~\textup{K}}$, and (c) the temperature for ${B=1~\text{T}}$ and ${I=10~\text{\textmu A}}$. Red dotted lines are linear fits. (d) Gate voltage dependence of $R_{l,0}$, $R_{l,\textup{max}}^{\textup{odd}}$ and ${Z\,R_{t,\textup{max}}^{\textup{odd}}}$, ${R_{\textup{UMR}}=R_{l,\textup{max}}^{\,\textup{odd}}-Z\, R_{t,\textup{max}}^{\,\textup{odd}}}$.}
\label{Fig3}
\end{center}
\end{figure}

We report in Fig.~\ref{Fig1}\textcolor{blue}{(b)} the 4-probe temperature dependence of the zero magnetic field resistance $R_{l,0}$. The resistance saturation at low temperature is a fingerprint of a conduction channel in parallel with the bulk (black dashed line) which we attribute to the presence of subsurface states. The angular dependence of $R_{l}^{\,\textup{even}}$ at  ${15~\textup{K}}$ in the $(xy)$ plane is shown in Fig.~\ref{Fig1}\textcolor{blue}{(c)} for ${I=10~\text{\textmu A}}$. This MR signal exhibits maxima (resp. minima) for ${\textbf{B}\parallel\hat{\textbf{y}}}$, $\varphi=90^{\circ}$ (resp.  ${\textbf{B}\parallel\hat{\textbf{x}}}$, $\varphi=0^{\circ}$). Since the sign is not reversed when reversing the magnetic field direction, we call this term anisotropic magnetoresistance (AMR) by analogy with ferromagnets. At  ${15~\textup{K}}$, we find an AMR of ${0.4\%}$ under a magnetic field of 1 T. The same behaviors are obtained for angular dependencies within $(zy)$ and $(zx)$ planes. \\
In Fig.~\ref{Fig2}, we report the angular dependence of $R_{l}^{\,\textup{odd}}$ and $R_{t}^{\,\textup{odd}}$ in the ($xy$), ($xz$) and ($yz$) planes for ${B=1~\text{T}}$, ${I=10~\text{\textmu A}}$ at ${15~\textup{K}}$. We observe a unidirectional behavior for both longitudinal and transverse resistances: the maximum (minimum) of $R_{l}^{\,\textup{odd}}$ is observed for ${\textbf{B}\parallel\varmp\hat{\textbf{y}}}$,  and the maximum (minimum) of $R_{t}^{\,\textup{odd}}$ is observed for ${\textbf{B}\parallel\varmp\hat{\textbf{x}}}$. Thus, experimentally, ${R_{l}^{\,\textup{odd}}=-R_{l,\textup{max}}^{\,\textup{odd}}\,\sin{(\varphi)}\sin{(\theta)}}$ and ${R_{t}^{\,\textup{odd}}= -R_{t,\textup{max}}^{\,\textup{odd}}\,\cos{(\varphi)}\sin{(\theta)}}$. These functions are shown as solid lines in Fig.~\ref{Fig2}. 
The angular dependence of the transverse resistance reveals the presence of the Nernst effect due to a current-induced vertical temperature gradient (along $\hat{\textbf{z}}$) in the Ge(111) film. This effect generates spurious thermal UMR signal in the longitudinal resistance. The Nernst effect contribution to $R_{l}^{\,\textup{odd}}$ can be written as: ${R_{l,\textup{max}}^{\,\textup{odd,Nernst}}=Z\, R_{t,\textup{max}}^{\,\textup{odd}}}$, with $Z$ being the aspect ratio of the channel (${Z=4}$ in our case) \cite{Avci2015}. Hence, to remove the Nernst effect contribution from the longitudinal signal, we study ${R_{\textup{UMR}}=R_{l,\textup{max}}^{\,\textup{odd}}-Z\, R_{t,\textup{max}}^{\,\textup{odd}}}$. We find that the Nernst effect contribution is negligible at ${15~\textup{K}}$ for low currents while it dominates when approaching room temperature and/or applying large currents (more details are given in the Supplementary Material).

In Fig.~\ref{Fig3} we investigate the dependencies of $R_{\textup{UMR}}$ on the applied current [Fig.~\ref{Fig3}\textcolor{blue}{(a)}], magnetic field [Fig.~\ref{Fig3}\textcolor{blue}{(b)}], temperature [Fig.~\ref{Fig3}\textcolor{blue}{(c)}] and gate voltage [Fig.~\ref{Fig3}\textcolor{blue}{(d)}]. The signal is normalized with respect to the zero field resistance $R_{{l},0}$ at the corresponding current. In agreement with previous reports on UMR generated by spin-momentum locking \cite{Olejnik2015,Avci2015} we observe a signal proportional to the current and the magnetic field. ${R_{\textup{UMR}}/R_{l,0}}$ is maximum and almost constant at low temperature ($T<{20~\text{K}}$) and sharply decreases when the temperature becomes comparable to the Rashba spin-splitting energy ($\approx60~\text{K}$). As shown in Fig.~\ref{Fig3}\textcolor{blue}{(d)}, the application of a top gate voltage modulates the channel resistance $R_{l,0}$. In Fig.~\ref{Fig3}\textcolor{blue}{(d)}, we also plot both the longitudinal and transverse odd resistance components as a function of the gate voltage. The transverse component we attribute to the Nernst effect stays constant with the gate voltage. This observation is consistent with the fact that this effect is due to vertical temperature gradient in the Ge(111) film and is almost unaffected by the top gate voltage. By contrast, the longitudinal component attributed to the UMR effect is much affected by the gate voltage: it increases from $V_g=-10~\text{V}$ to $V_g=+10~\text{V}$ by a factor $\approx3$. $R_{\textup{UMR}}$ cancels out at $V_g=-10~\text{V}$ and increases from $-10~\text{V}$ to $+10~\text{V}$.

\begin{figure}[t!]
\includegraphics[width=0.5\textwidth]{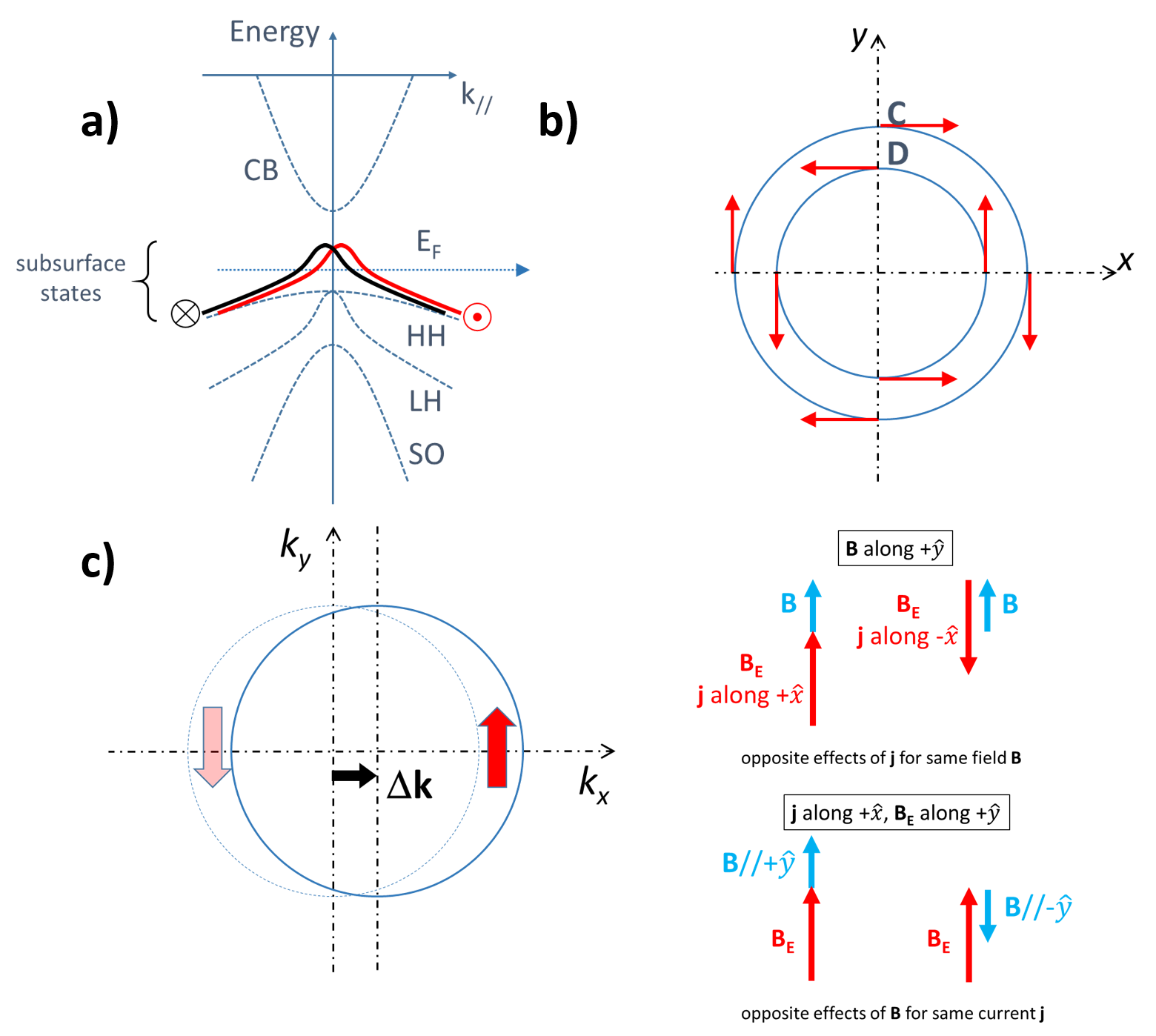} 
\caption{(color online) (a) Schematics of the Ge(111) electronic band structure (in agreement with Ref.~\onlinecite{Ohtsubo2013}) showing the bulk conduction and valence bands. The Fermi level is at a position corresponding to a $p$-doped film. Subsurface states are located just above the maximum of the bulk valence band and are crossed by the Fermi level. They are spin-splitted by the Rashba and atomic spin-orbit interactions. (b) Fermi contours of the subsurface states. The outer (inner) contour is named C (D) with clockwise (counter-clockwise) spin helicity. (c) Illustration of the combined effects of the applied magnetic field $\textbf{B}$ and the current dependent pseudo-magnetic field $\textbf{B}_{\textup{E}}$ on the resistivity of subsurface states for a single contour (D here). The contour is shifted by $+~\Delta\textbf{k}$ due to the application of a current density $\textbf{j}$ along $+~\hat{\textbf{x}}$. The current direction and spin helicity set the pseudo-magnetic field $\textbf{B}_{\textup{E}}$.}
\label{Fig4}
\end{figure}

To make a comparison with previous results on different systems, we can define a figure of merit $\eta$. Since the UMR signal is proportional to the current and magnetic field, a natural definition is: ${\eta=R_{\textup{UMR}}/(R_{l,0}\,j\,B)}$. At ${15~\text{K}}$, in Ge(111), we obtain ${\eta=2.5\times10^{-4}~\textup{cm}^{2}/(\textup{A\,T})}$ when considering the charge current flowing in the whole Ge(111) film (${2~\text{\textmu m}}$) and ${\eta=4.2\times10^{-7}~\textup{cm}^{2}/(\textup{A\,T})}$ if we consider that the current completely flows within the spatial extension of the subsurface states (10 atomic layers from Ref.~\onlinecite{Aruga2015}). In the worst case scenario, the value of $\eta$ obtained in Ge(111) is orders of magnitude larger than the one of SrTiO$_3$ at 7 K [${\eta=2\times10^{-9}~\textup{cm}^{2}/(\textup{A\,T})}$ from Ref.~\onlinecite{He2018}] and the one of Bi$_2$Se$_3$ at ${60~\text{K}}$ [${\eta=2\times10^{-11}~\textup{cm}^{2}/(\textup{A\,T})}$ from Ref.~\onlinecite{He2017}]. In this second case, if we compare the results extrapolated at ${58~\text{K}}$ for Ge(111) we still obtain a larger $\eta$ value [${3\times10^{-8}~\textup{cm}^{2}/(\textup{A\,T})}$]. 

At variance with previously reported systems \cite{He2017,He2018} the UMR is isotropic with respect to the direction of the current flow in the surface Brillouin zone (SBZ). In fact, in the data shown in Figs.~\ref{Fig1}\textcolor{blue}{$-$}\ref{Fig3}, the current flows along the $\Gamma\text{M}$ direction of the Ge(111) SBZ, but no difference, within the experimental error, is detected with the current flowing along other reciprocal lattice directions (see Supplementary Material). In Refs.~\onlinecite{He2017,He2018}, the magnetoresistance is affected by the direction of the current flow in the SBZ, indicating that, in such a case, the UMR originates from the out-of-plane spin texture. In the case of Ge, this contribution appears to be negligible. We thus propose an alternative mechanism, in which the UMR in Ge(111), results from a combination of the applied magnetic field and the current-induced pseudo-magnetic field in the spin-splitted subsurface states of Ge(111) shown in Fig.~\ref{Fig4}\textcolor{blue}{(a)}. Ge(111) subsurface states are located close to the top of the valence bands and can only contribute to transport in $p$-type Ge(111) \cite{Ohtsubo2013}. This interpretation is supported by the fact  that we do not observe this effect for $n$-type Ge(111) (see Supplementary Material). It also explains the gate voltage dependence of $R_{\textup{UMR}}$ in Fig.~\ref{Fig3}\textcolor{blue}{(d)}. Applying negative gate voltage shifts the Fermi level down into the valence band which leads to the activation of bulk conduction and $R_{\textup{UMR}}\approx0~\Omega$ for $V_g=-10~\text{V}$. At variance, by ramping the gate voltage from ${-10~\text{V}}$ to ${+10~\text{V}}$, the Fermi level shifts into the subsurface states thus increasing $R_{\textup{UMR}}$. Finally, this interpretation also explains the temperature dependence of the UMR. By increasing the temperature, bulk conduction in the valence band is activated and shorts the subsurface states. Moreover, the Rashba spin-orbit coupling of $\sim58\,k_B$ in Ge subsurface states \cite{Ohtsubo2013} becomes negligible with respect to $k_{\textup{B}}T$ suppressing spin-momentum locking.\\
For the Fermi level crossing the subsurface states as shown in Fig.~\ref{Fig4}\textcolor{blue}{(a)}, the Fermi contour is made of two concentric rings  [C and D in Fig.~\ref{Fig4}\textcolor{blue}{(b)}]  with opposite spin helicities. To describe the magnetotransport inside the subsurface states, we consider the following model Hamiltonian :
\begin{equation} \label{eq1}
\mathscr{H}=-\frac{\hbar^{2}k^{2}}{2m^{*}}+\alpha\,(\textbf{k}\times\bm{\upsigma})\cdot\hat{\textbf{z}}+g\mu_{\textup{B}}\bm{\upsigma}\cdot\textbf{B},
\end{equation}
with $\hbar$ being the reduced Planck constant, $m^{*}$ the effective mass of holes in the subsurface states, $\alpha$ the Rashba spin-orbit interaction, $\bm{\upsigma}$ the vector of Pauli matrices, $g$ the Land\'e factor and $\mu_{\textup{B}}$ the Bohr magneton. 
When a 2D charge current density $\textbf{j}$ flows in the subsurface states, in the Boltzmann approach, the momentum acquires an extra component ${\Delta\textbf{k}=\beta\textbf{j}}$ with ${\beta=4\pi/(e v_{\textup{F}}k_{\textup{F}})}$, $v_{\textup{F}}$ and $k_{\textup{F}}$ the Fermi velocity and wavevector we consider ($e=\lvert e \rvert$). A well-known consequence of such shifts of Rashba Fermi contours is the Rashba-Edelstein spin polarization \cite{Chappert2007} due to the unbalance between the opposite spin polarizations induced by the shifts in the same direction of the Rashba-splitted Fermi contours of opposite helicity. In parallel with the Rashba-Edelstein effect, the shift $\Delta\textbf{k}$ introduces  a current-induced out-of-equilibrium energy term which, from Eq.~\ref{eq1}, is equal to $\alpha(\Delta\textbf{k}\times\bm{\upsigma})\cdot\hat{\textbf{z}}=\alpha\beta(\textbf{z}\times\textbf{j})\cdot\bm{\upsigma}$ and acts on the spins as a pseudo-magnetic field  ${\textbf{B}_{\textup{E}}=(\alpha\beta/g\mu_{\textup{B}})\,\hat{\textbf{z}}\times \textbf{j}}$. As illustrated in Fig.~\ref{Fig4}\textcolor{blue}{(c)}, for a current along $\pm~\hat{\textbf{x}}$ with $\alpha > 0$ , this field is directed along $\pm~\hat{\textbf{y}}$ and proportional to the current density. In the presence of an applied magnetic field $\textbf{B}$, the spin of the subsurface states is submitted to $\textbf{B}+\textbf{B}_{\textup{E}}$, $\textbf{B}_{\textup{E}}$ increasing or decreasing the effect of the $y$ component of $\textbf{B}$ for currents either along $\textup{+}$ or $\textup{-}$ $\hat{\textbf{x}}$.
In the same way, still for $\alpha > 0$ for $\textbf{j}$ along $+\hat{\textbf{x}}$ and $\textbf{B}_{\textup{E}}$ along $\hat{\textbf{y}}$ , there is addition or subtraction of the effects of $\textbf{B}$ and $\textbf{B}_{\textup{E}}$ for opposite orientations of  $\textbf{B}$ along $\hat{\textbf{y}}$. The physics of the UMR thus comes from the pseudo-field $\textbf{B}_{\textup{E}}$ induced by the out-of-equilibrium situation of a current flow and acting on the spins.  
We can go a little further by assuming that the AMR term shown in Fig.~\ref{Fig1}\textcolor{blue}{(c)} (the only MR in the limit $j \rightarrow 0$) is also due to the effect of $\textbf{B}$ on the spins. We thus follow Taskin $\textit{et al.}$ \cite{Taskin2017} who explain the AMR of Rashba systems by the re-introduction of some backscatterings by a partial re-alignement of the spins by $\textbf{B}$ and we neglect contributions such as the effect of the Lorentz force on the trajectories. 
Then, in the situation of finite $j$, we add $\textbf{B}_{\textup{E}}$ to $\textbf{B}$ in the $B^2$ term of the AMR to derive the expression of UMR. The AMR term can be written as: 
\begin{equation} \label{eq2}
{\left(\Delta R/R\right)_{\textup{AMR}}=-A\,B^2\,\cos^2{(\varphi)}=AB_{y}^2-A\,B^2}\,
\end{equation}
Where ${A\approx0.004}$. Adding $B_{\textup{E}y}=\alpha\beta j/g\mu_{\textup{B}}$ to $B_\textup{y}$, and keeping only the terms of first order in $j$ gives :
\begin{equation} \label{eq3}
\Delta R/R=-AB^2\textup{cos}^2(\varphi)+2A(\alpha\beta /g\mu_{\textup{B}})\,j\,B\,\textup{sin}(\varphi)\,
\end{equation}
Where the second term, proportional to $j\,B$, is the UMR.  Our experimental results with an UMR proportional to $j\,B\,\textup{sin}(\varphi)$, see [Fig.~\ref{Fig2}], correspond to a negative value of the Rashba coefficient $\alpha$, that is to the clockwise chirality of the spin orientation in the outer Fermi contour. This chirality is in agreement with the chirality derived from spin-resolved ARPES measurements for the subsurface states inside Ge at Ge/Bi interfaces, as shown in Fig. 3a of [\onlinecite{Ohtsubo2010}]. 
Quantatively, taking reasonable values for the parameters in the expression of the UMR amplitude. 
By setting ${B=1~\text{T}}$, ${j=0.33~\textup{A}\,\textup{m}^{-1}}$ in the subsurface states, ${\alpha=-0.2~\textup{eV}\cdot\textup{\AA}}$ (in \cite{Ohtsubo2010}, this value corresponds to Bi covered subsurface states, in our case it is probably an upper bound), ${k_{\textup{F}}=0.025~\textup{\AA}^{-1}}$ (Rashba splitting $\lvert\alpha k_{\textup{F}}\rvert=5$ meV$\sim58\,k_B$), $m^{*}=0.4\,m_e$ \cite{ioffe}, $m_e$ being the electron mass, ${v_{\textup{F}}=\hbar k_{\textup{F}}/m^{*}}$ and ${g=2}$, we find a UMR amplitude of ${\approx0.2\%}$. This value is in good agreement with our low temperature experimental data. 
We indeed find a maximum value of $0.5\%$ at  ${15~\textup{K}}$. Therefore, by using simple arguments, we capture the physics of UMR in the Ge Rashba-splitted subsurface states.

In conclusion, we performed magnetoresistance measurements on Ge(111) and detected a unidirectional magnetoresistance (UMR)  which  scales  linearly  with  both  the  current  and  the  applied  magnetic  field. We ascribe the UMR to the spin-momentum locking generated by the Rashba effect in the subsurface states of Ge(111) and interpret our results in a simple model relating the UMR to the Rashba coefficient and the characteristic parameters of the subsurface states. Such unidirectional effects can be expected in any Rashba 2DEG and can be used to obtain information about the electronic structure details. The  amplitude of  the  detected  UMR signal  is  much  larger  than  the  ones  previously  reported. We  also  showed  that  this  UMR  is tunable  by  turning  on  and  off  the  Rashba  coupling  in  the  conduction  channel  by applying  a  gate  voltage. Ultimately, these findings lead towards the development of a semiconductor-based spin transistor where the spin information can be manipulated by a gate-tunable Rashba field.\\
The authors acknowledge the financial support from the ANR project ANR-16-CE24-0017 TOPRISE. One of us (AF) acknowledges fruitful discussions with A .Dyrdal and J. Barnas (Poznan University), as well as with S. Zhang (University of Arizona). 

\end{sloppypar}

\end{document}